\documentclass[preprint]{aastex63}

\received{}
\revised{}
\accepted{}
\submitjournal{AJ}

\shorttitle{VY CMa ALMA Clumps} 
\shortauthors{Humphreys et al.}

\begin{document}

\title{The Hidden Clumps in VY CMa Uncovered by ALMA}

\correspondingauthor{Roberta Humphreys}
\email{roberta@umn.edu}

\author{Roberta M. Humphreys}
\affiliation{Minnesota Institute for Astrophysics,  University of Minnesota, Minneapolis, MN 55455, USA}

\author{A. M. S. Richards}
\affiliation{Department of Astronomy University of Manchester, UK}

\author{Kris Davidson}
\affiliation{Minnesota Institute for Astrophysics,  University of Minnesota, Minneapolis, MN 55455, USA}

\author{A. P. Singh}
\affiliation{Department of Chemistry and Biochemistry University of Arizona, USA}

\author{L. Decin}
\affiliation{Department of Physics and Astronomy, KU, Leuven, Belgium}

\author{L. M. Ziurys}
\affiliation{Department of Astronomy; Department of Chemistry, and Steward Observatory University of Arizona, USA}

\begin{abstract}
	The red hypergiant VY CMa is famous for its very visible record of high mass loss events. Recent CO observations with ALMA revealed three previously unknown large scale outflows (Paper I). In this paper we use the CO maps to investigate the motions of a cluster of four clumps close to the star, not visible in the optical or infrared images. We present their proper motions measured from two epochs of ALMA images and determine the line of sight velocities of the gas in emission at the clumps. We estimate  their masses and ages, or time since ejection, and conclude that all four were ejected during VY CMa's active period in the early 20th century. Together with two additional knots observed with HST, VY CMa experienced at least six massive outflows during a 30 year period with a total mass lost $\ge$ 0.07 M$_{\odot}$. 
The position-velocity map of the $^{12}$CO emission reveals previously 
unnoticed attributes of the older outer ejecta. In a very 
 narrow range of Doppler velocities, $^{12}$CO absorption and emission 
 causes some of this outer material to be quite opaque.  At those 
 frequencies the inner structure is hidden and we see only emission 
 from an extended outer region. This fact produces a 
 conspicuous but illusory dark spot if one attempts to subtract the continuum  
 in a normal way.  
\end{abstract}


\section{VY CMa's Mass Loss History and the ALMA Clumps} \label{sec:intro}

Mass loss is observed across the upper HR Diagram and alters the evolution 
of the most massive stars. It may be slow and continuous or occur in more 
dramatic high mass loss events which may be irregular or  single giant eruptions. In the red and yellow supergiants, these mass loss events may determine their final fate whether as supernovae or direct collapse to the black hole. 

\vspace{2mm} 

The red hypergiant VY CMa is one of the most important evolved massive stars for understanding the role of high mass  loss episodes on the final stages of the majority of massive stars which pass through the red supergiant stage.  Its mass loss history is  highly visible in optical and infrared images with  prominent arcs and filaments and clumps of knots throughout its extended diffuse ejecta. Doppler velocities of  K I emission, together with proper motions of the  discrete ejecta \citep{RMH2005,RMH2007,RMH2019} demonstrated that the arcs and clumps of knots are spatially and kinematically distinct from the diffuse ejecta and were expelled in separate events over several hundred years with several episodes in the last century \citep{RMH2021}

\vspace{2mm}

New high resolution images with ALMA \citep{Singh}, hereafter Paper I, reveal a more extended diffuse ejecta with at least three previously unknown prominent large arcs or outflows observed in CO and HCN emission. The addition of three major outflows expands our record of its high mass loss episodes and frequency, and presents  additional challenges to possible models for their origin in VY CMa and similar luminous red supergiants.

\vspace{2mm} 

 ALMA Science Verification sub-mm and  continuum emission images of VY CMa at 321, 325 GHz and 658 GHz \citep{Richards,OGorman} earlier revealed an additional, large, potentially massive clump or knot close to the star.  It is  not visible  in the  optical HST images.  Designated Clump C  by  \citet{Richards}, it is optically thick at these frequencies with an estimated dust mass of 2.5 $\times$ 10$^{-4}$ M$_{\odot}$, or 5 $\times$ 10$^{-2}$ M$_{\odot}$ assuming a gas to dust ratio of 200. Using the Science verification data for Band 5 at 163 - 211 GHz, \citet{Vlemmings} concluded  that the dust in Clump C was optically thick even at 178 GHz, and estimated a dust mass of $>$ 1.3 $\times$ 10$^{-3}$ M$_{\odot}$ or a total mass (dust + gas) of more than 0.1 M$_{\odot}$.  Making it potentially the most massive ejecta in VY CMa. 

\vspace{2mm}

It is not surprising that Clump C has increased questions about VY CMa's mass loss history and its mass loss mechanism.  Using additional ALMA observations in Bands 6 and 7 with a higher angular resolution of 0$\farcs$1, \citet{Kaminski} identified three additional features or smaller clumps clustered near Clump C  and the star. See Figure 3 in \citet{Kaminski} and our Figure 1. All are dusty and he comments that the dust mass for Clump C may be even higher than suggested by \citet{Vlemmings}, but notes that porosity would allow a lower dust mass. For example, with higher resolution ALMA images, \citet{Asaki} show that Clump C is resolved into many small condensations  with a range of sizes  and intensities.

\vspace{2mm}  
Despite its significance, there is much we don't know about Clump C and associated knots such as their motions, spatial orientation with respect to VY CMa,  and when they were ejected. Do they represent a single massive eruption or are they from separate events over a period of time?   In this paper we report on the proper motions of Clump C and its neighbors measured from two epochs of continuum images observed with ALMA, and the motions of the CO gas in emission at the clumps.  Hereafter, to distinguish them from the numerous knots, clumps etc observed in the optical images, we refer to them in this paper as the ALMA clumps.

\vspace{2mm} 

In the next section we describe our observations and proper motion measurements. In section three  we discuss the line of sight velocities of the CO emission at the four ALMA clumps. Section 4  summarizes their total motion, orientation and ages or time since ejection and relation to VY CMa's other recent mass loss events.  In section 5, we  estimate the masses of the clumps and review the energies involved in these high mass loss events. The outer ejecta revealed by the  CO emission and its obscuration at  Clump C is discussed in Section 6. The final  section is a review of questions about VY CMa's evolutionary state.        

\vspace{2mm} 

\section{Data  and  Proper Motion Measurements }  

Our high resolution observations\footnote{project 2021.1.01054.S}, obtained with ALMA in Band 6 covering frequencies 216 to 270 GHz, are described in Paper I.  In this paper we use the continuum image at 249 GHz and the CO image cube with spatial resolution 0\farcs2  and velocity resolution of 1.25 km s$^{-1}$. The observational parameters are described in Table A1 in the Appendix.
Our continuum image at 249 GHz is shown in Figure 1. Figure 2 shows the $^{12}$CO emission at the clumps in two representative channels. The feature labeled B$^{\prime}$ is an outflow not mentioned in previous papers, and is not present in the continuum images \footnote{B$^{\prime}$ is observed in the $^{12}$CO data at Doppler velocities 
$+31 \lesssim V_{LSR} \lesssim +47$ km s$^{-1}$, 
but not in the continuum (see Figure 1).  Hence we cannot estimate its 
proper motion.  Lacking appreciable continuum emission, most likely it 
has less mass than A, B, C, or D.}.     Also note that the brightest $^{12}$CO emission is not centered on the position of VY CMa. This apparent offset to the east is observed at all the frequencies in our image cubes with bright CO emission from the star's circumstellar ejecta. The peak of the $^{12}$CO emission line profile at VY CMa  shown in the Appendix,  has a blue shift with respect to the star's expected LSR velocity of 22 km s$^{-1}$. Thus a cloud of  CO gas (and dust) may be asymmetric and  probably expanding relative to VY CMa.  

\vspace{2mm}

The data reduction steps and calibration are described in Paper I and its Appendix. A descriptive summary is in Appendix A.

\begin{figure}
\figurenum{1}
\epsscale{0.8}
\plotone{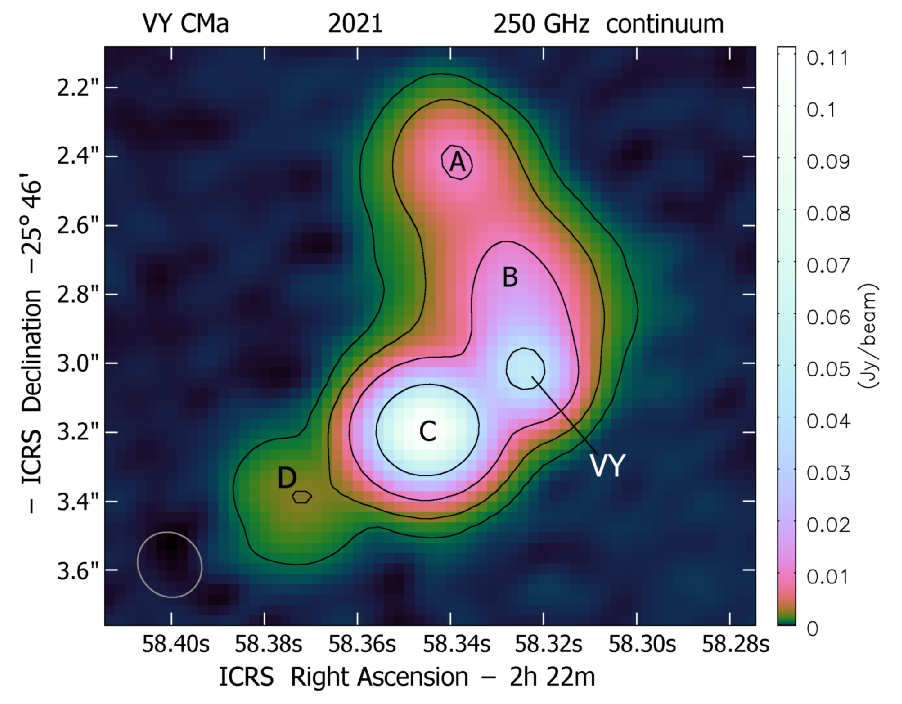}
	\caption{The continuum image at 249 GHz. The contours showing the structure and outlines of the clumps  are in multiples of  (1, 4, 16, 64) $\times \, I_1$,  where  $I_1 \approx$ 0.6 mJy/beam $\approx$ \, 8 mJy arcsec$^{-2}$.   The brightness temperature at the outer contour is about 0.2 K. The synthesized beam (FWHM) is shown as an ellipse in the lower left corner.}  
\end{figure}

\begin{figure}
\figurenum{2}
\epsscale{0.8}  
\plottwo{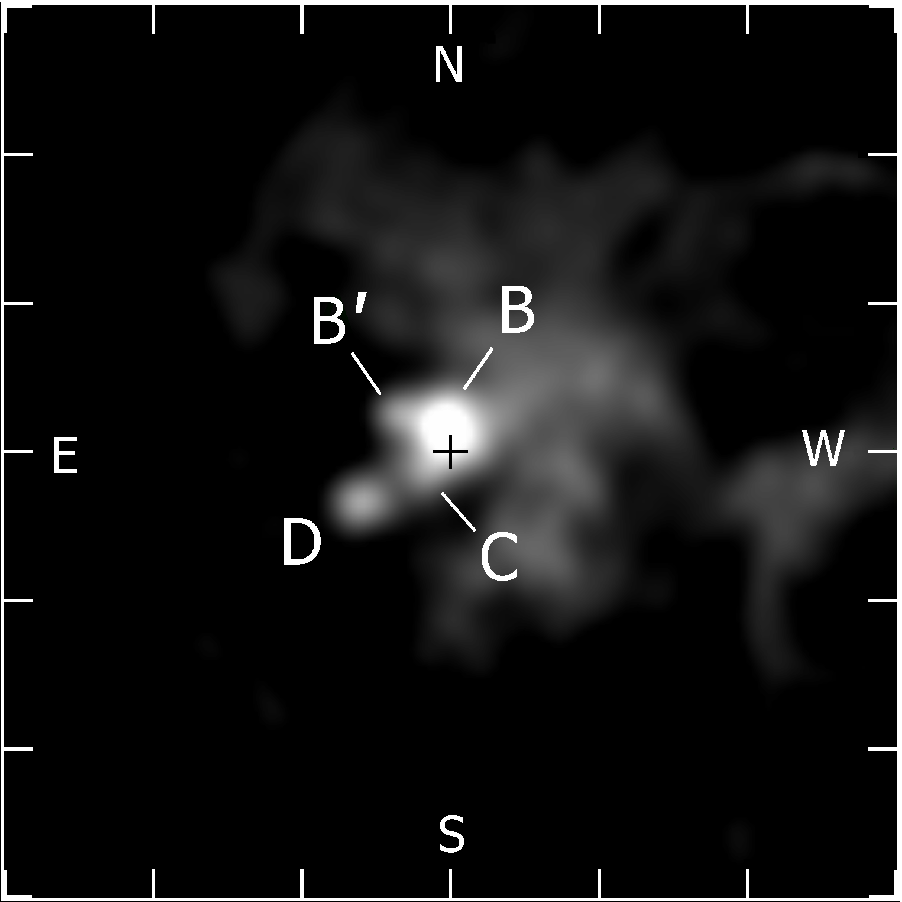}{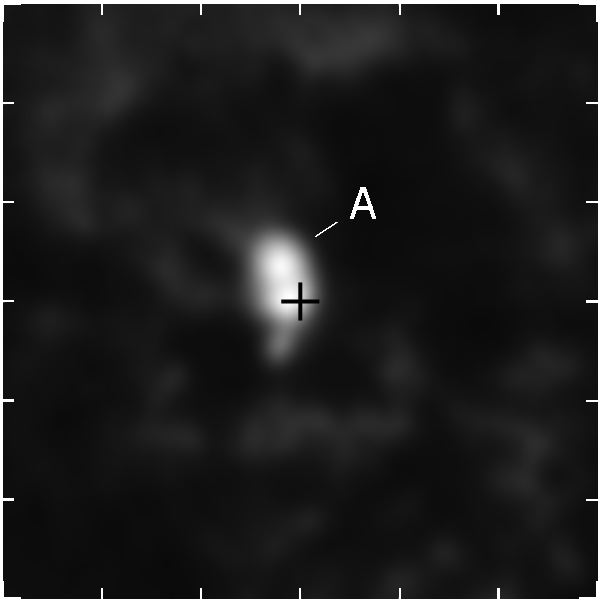}
	\caption{$^{12}$CO images of a $6\arcsec \times 6\arcsec$ region centered on VY CMa.  The tick marks are $1\arcsec$. Clumps C and D are redshifted relative to VY CMa and observed over the same range of Doppler velocities while Clump A is blueshifted.  Clump B is not resolved from the bright CO emission from VY CMa's circumstellar ejecta. (a) $^{12}$CO emission at V$_{LSR}$ 43 km s$^{-1}$ toward Clumps C and D, and new outflow B$^{\prime}$. The position of Clump B is marked. (b) toward Clump A at V$_{LSR}$ +6 km s$^{-1}$.  The position of VY CMa is shown as a $+$ in each panel; its Doppler velocity is $+22$ km s$^{-1}$ relative to the LSR.  The brightness scales 
differ between panels. The images are continuum substracted.}
\end{figure}

\subsection{Proper Motions  of the Clumps }

To measure the proper motion of Clump C and the other knots, we compare their  positions relative to VY CMa in our  continuum image from 2021 December 09 with an earlier image.  The closest dataset to ours in terms of frequency and resolution with a decent time scale is from Kaminski, project 2013.1.00156.S from 2015 September 27 giving a 6.2 yr baseline. The continuum images used for the proper motions, the measurement method, and   uncertainties are described in the Appendix.  

\vspace{2mm} 
The separate measurements for 2015 and 2021 are given in Table 1 for the four clumps. Their proper motions, positions, and derived parameters are summarized in Table 2.  For example, for Clump C, the continuum images yield separations of  314 mas in 2015 and  341 mas in 2021 with respect to VY CMa.   The increase in angular separation of 23 $\pm$ 2 mas at a distance of 1.15 kpc \footnote{The weighted average of \citet{Choi}(1.14 (+0.11 -0.09) kpc) and \citet{Zhang}(1.20 (+0.13 -0.10) kpc) with a combined uncertainty of $\approx \pm$ 0.08 kpc.}, is 26.5$\pm$ 2.8 AU. In 6.2 yrs, the proper motion is 3.7 mas per year and the  transverse velocity in the plane of the sky is thus 20.2 $\pm$ 2.3 km s$^{-1}$.    The angular change in Declination and Right Ascension between 2015 and 2021 gives the direction of motion ($\phi$) for Clump C, 102$^{\circ}$,  measured N through E.  The same parameters are included in Table 2  for the other clumps.  

\vspace{2mm} 
Figure 3 is a map showing the measured proper motions and the direction the clumps are moving.  Clumps C and B have a direction of motion from their proper motions that do not point radially back to the star. The uncertainties in $\phi$, the direction of motion in Table 2 however, are within  three sigma of the position angle for Clumps C and B. Other factors may influence the direction the clumps appear to be moving such as the overall asymmetry of the circumstellar ejecta near VY CMa and the role of magnetic fields, see Section 5.

\begin{deluxetable}{llrrclrrc} 
\tablewidth{0 pt}
\tabletypesize{\scriptsize}
\tablenum{1} 
\tablecaption{Proper Motion Measurements}
\tablehead{
\colhead{Date} & 
\colhead{Object} &
\colhead{R.A.} &
\colhead{Dec} & 
\colhead{Total Offset} &   
\colhead{Date} &
\colhead{R.A.} &
\colhead{Dec} & 
\colhead{Total Offset} 
}
\startdata
2015.74 & VY CMa & 7:22:58.32437  & -25:46:3.0687 & \nodata &  2021.94  &  7:22:58.32365 & -25:46:3.0148 & \nodata \\
        &        &                &               &         &           &                &               &          \\
  "	&  Clump A  &    58.33704  &       2.5366 & \nodata &   "       &       58.33817 &        2.4350 &  \nodata  \\
	&  A-VY   &      0.01267E &        0.5321N & \nodata &         &        0.01452E  &        0.5798N & \nodata    \\
	&  offset mas  &  172E mas &      532N mas &  559$\pm$11.8 mas &  &    197E mas  &   580N mas &   612mas$\pm$4.1   \\  
	&             &             &             &          &   &                &               &       \\           	 
   "	&  Clump B  &    58.32545  &      2.8422  & \nodata   &   "    &        58.32665 &       2.7704 &  \nodata     \\
	 & B-VY    &     0.00108E  &     0.2265N  &  \nodata  &        &       0.00300E   &        0.2444N & \nodata     \\
	 & offset mas &    15E mas   &   226N  mas & 226$\pm$8.9 mas  &    &   41E mas  &     244N mas  &  247$\pm$3.4 mas \\
       	&             &             &             &          &   &                &               &       \\
  "	& Clump C     &   58.34395  &     3.2428  &  \nodata &   "      &       58.34498  &       3.1938  &  \nodata  \\
	& C-VY        &  0.01958E    &    -0.1741S  &  \nodata &         &      0.02133E  &      -0.1790S & \nodata    \\
	& offset mas  &   266E mas   &    174S mas   &  318$\pm$1.8 mas &    &   290E mas &    179S mas  &  341$\pm$0.9 mas \\
         &             &             &             &          &   &                &               &       \\
  " 	& Clump D     & 58.36965     &     3.4235  & \nodata  &    "       &    58.37204   &      3.3897  & \nodata  \\
	& D- VY       &   0.04528E   &    -0.3282S &  \nodata &            &   0.04838E    &     -0.3749S & \nodata   \\
	& offset mas  &   616E mas    &    328S mas & 698$\pm$23.3 mas &  &   658E mas    &    375S mas  &  757$\pm$10.0 mas  
\enddata
\end{deluxetable} 

\begin{deluxetable}{lllllll} 
\tablewidth{0 pt}
\tabletypesize{\footnotesize}
\tablenum{2} 
\tablecaption{Measured Positions and Proper Motions}
\tablehead{
\colhead{Clump} & 
	\colhead{Proj Dist\tablenotemark{a}}  &
\colhead{Pos. Angle}  &  
	\colhead{Angular Sep.\tablenotemark{b}} &
\colhead{Proper Motion} &
\colhead{Trans. Vel.} & 
\colhead{Direct. $\phi$} \\
\colhead{}   &
\colhead{mas} &  
\colhead{degrees} &  
\colhead{mas}  & 
\colhead{mas yr$^{-1}$} &
\colhead{km s$^{-1}$} &
\colhead{degrees} 
}
\startdata
	Clump A & 560  &  19  & 53$\pm$12 & 8.5$\pm$2.0 & 46$\pm$11 & 27.5 $\pm$ 13  \\
	Clump B & 233   &  9    &  21$\pm$9.5  &  3.4$\pm$1.5 & 18$\pm$9  & 55.3 $\pm$ 17  \\ 
	Clump C & 320  &  122 &  23$\pm$2.0 & 3.7$\pm$0.4 &  20$\pm$2    & 102.0 $\pm$ 5 \\
	Clump D  & 716  & 120 &   59$\pm$25.3  & 9.5$\pm$4.1 & 52$\pm$23    &  138.3 $\pm$ 29  
\enddata
	\tablenotetext{a}{The projected distance from VY CMa in mas, is the average of the separation from \citet{Kaminski} and for 2015 in Table 1.} 
	\tablenotetext{b}{The increase in angular sparation or distance from VY CMa from 2015 to 2021.} 
\end{deluxetable} 

\begin{figure}  
\figurenum{3}
\epsscale{0.5}
\plotone{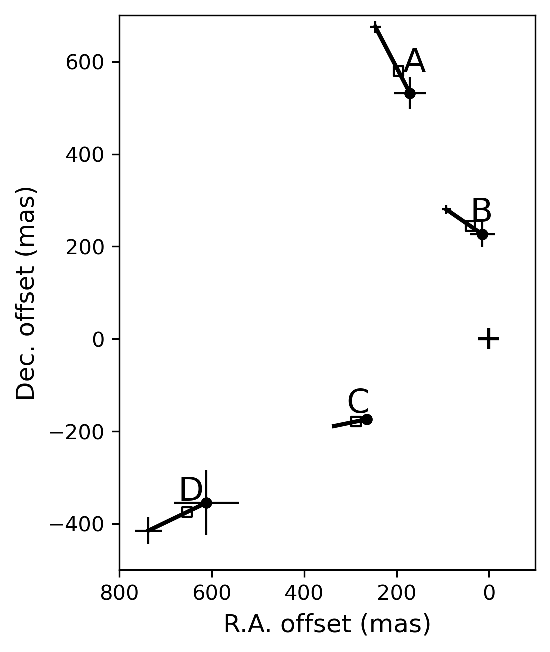}
\caption{Map of the proper motions showing the direction of motion for the clumps relative to their positions in 2015.  The 2015 positions are marked by solid circles and the 2021 positions by hollow squares. The vectors and error bars have been multiplied by three to make them more visible. } 
\end{figure}

\section{Line of Sight Velocities Toward the ALMA Clumps} 

At 200 mas resolution, the CO emission in this region is partially resolved into small condensations or features of a size similar to or slightly larger than the beam.  Figure 2 shows that  Clumps C and B are not well separated from the bright $^{12}$CO emission from VY CMa's circumstellar ejecta.  Clump D, however, appears as a small, emission peak or spot and is spatially separate from the circumstellar  $^{12}$CO emission in several channels. Clump A is also visible in several blueshifted channels as a relatively bright CO emission spot marginally separate from VY CMa.  We find that C and D are optically thick in the continuum but A and B are optically thin (Section 5).  As discussed below, clumps C and D together mark an interesting continuous structure or outflow. 

\vspace{2mm} 

In Table 3 we summarize the range of Doppler velocities measured   toward  each clump in the $^{12}$CO and $^{13}$CO image  cubes.  The velocities relative to the LSR are measured with respect to their respective rest frequencies. VY CMa has an LSR  velocity of 22$\pm$1 km s$^{-1}$  based on OH and SiO maser emission from numerous sources.  The velocities toward the ALMA clumps demonstrate  that clumps C and D are redshifted, moving away from and projected behind VY CMa while clump A is blueshifted relative to the star. Clump B is more uncertain.  \citet{Kaminski} reported a wider range of velocities from molecular emission from the clumps which can be seen in the line profiles in the Appendix, but  Figure 5 discussed below shows  that the higher positive and negative velocities are from extended and diffuse separate features.

\begin{deluxetable}{lll} 
\tablewidth{0 pt}
\tabletypesize{\small}
\tablenum{3} 
\tablecaption{V$_{LSR}$ Range\tablenotemark{a} for the CO  Emission Observed at the Clumps}  
\tablehead{
\colhead{Clump} &  
\colhead{$^{12}$CO em} &
\colhead{$^{13}$CO em}  \\ 
\colhead{}	     &
	 \colhead{km s$^{-1}$}  & 
	    \colhead{km s$^{-1}$}   
}
\startdata
Clump A &  +15 to -4  & +15 to -11   \\  
Clump B &  +30 to +43   & +30 to +44    \\
Clump C &  +32  to +48   & +30 to +44   \\
	Clump D & +27 to +48  & +26 to +47   
\enddata
	\tablenotetext{a}{VY CMa has V$_{LSR}$ of +22$\pm$ 1 km s$^{-1}$.} 
\end{deluxetable} 

\vspace{2mm} 

Although the CO emission at Clump D appears to be spatially separate from the emission associated with Clump C and   the stellar envelope in Figure 2,  the continuum image suggests an outflow to the southeast of the star with  D at its outer extremity. Figure 4 shows the summed $^{12}$CO image over 28 frequency channels covering the range in V$_{LSR}$ from +49 to +15 km s$^{-1}$. It clearly illustrates the  continous emission structure to the SE of the star, although the separate channels show that  C and D represent  different parts of this elongated feature.  

\vspace{2mm} 

In each outflow event there is usually a strong correlation between the Doppler velocity and distance from the star. Hence a position-velocity slice at an  oblique angle in the image cube can reveal the local morphology discussed here. 
Figure 4 illustrates the procedure. The  relevant plane in the image cube is defined by position angle 120$^{\circ}$ which samples both C and D with our spatial resolution. At each frequency channel, we measure the intensity of the signal at positions along a line or slit at 120$^{\circ} \pm 0.05$ arcsec. The extractions for each frequency are stacked to produce a 2-dimensional map where one axis is spatial and the other is Doppler velocity. The resulting position-velocity map is shown in Figure 5 for the LSR velocity range -20  to +80 km s$^{-1}$ and reveals the morphology of $^{12}$CO emission associated with Clumps D and C.

\vspace{2mm} 

Figure 5 also reveals  bands of whispy  or ``cirrus-like''  emission stretching horizontally across the lower half and top of the figure. Based on their brightness and size scale the ``cirrus'' bands  represent older ejecta from VY CMa moving at $\approx$ $\pm$  30 -- 40 km s$^{-1}$ relative to the star.  A prominent emission feature is also visible  nearly aligned in position with the brighter star at $\approx$ +30 to +40  km s$^{-1}$, at $r \lesssim 400$ AU.  The dark band at about -3 km s$^{-1}$ that appears to cut off the extended CO emission, is due to optically thick outer ejecta discussed in Section 6.

\begin{figure}
\figurenum{4}
\epsscale{0.5}
\plotone{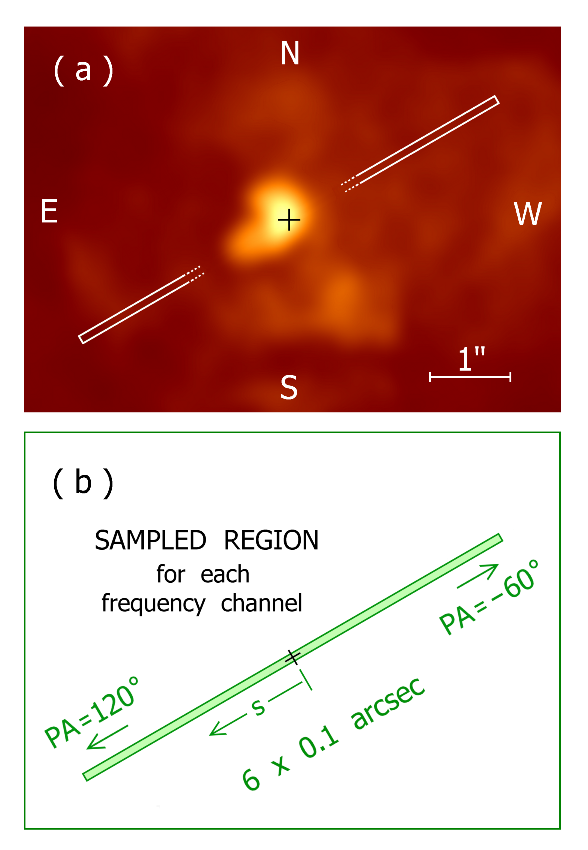}
\caption{Sketch illustrating how Fig.5 was produced. The  upper panel is a sum of the  $^{12}$CO images at Doppler velocities from $+15$ to $+49$ km s$^{-1}$ 
  relative to the LSR, and shows a virtual sampling
	slit oriented at position angle 120$^{\circ}$ to include Clumps C and 
  D (Fig. 2).  The lower panel shows its spatial parameters.  
  Intensity along that virtual slit was measured for each frequency 
  channel, and the resulting 1-D data sets were stacked to produce 
  the 2-D Fig. 5. }
\end{figure}

\begin{figure}
\figurenum{5}
\epsscale{1.2}
\plotone{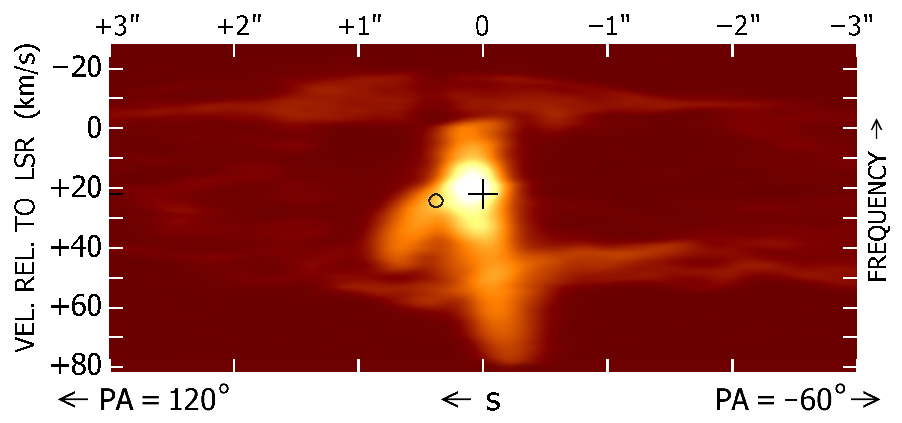}
	\caption{Position-Velocity map of the continuum subtracted $^{12}$CO emission at position angle 120$^{\circ}$, cf, Fig. 4. 
The vertical coordinate is the Dopper velocity, while the 
  horizontal scale is the spatial position along a line oriented  
	at position angle $120^{\circ}$ (roughly ESE).  The  
  location of the star is marked `$+$, and the approximate postion of 
	Clump C by a small circle.  To some extent the 
  Doppler velocities are correlated with location along our 
  line of sight, with the `near side' at the top of the figure.   
  However, at least two sets of ejecta with different ages 
  and size scales are superimposed, e.g.\ at velocities 
  +30 to +50 km $^{-1}$. Also see Fig. 8.  }
\end{figure}

\subsection{Outflow Velocities of the Clumps}   

The prominent  diffuse emission feature from V$_{LSR}$ +45 to about +25 km s$^{-1}$ in Figure 5 extends from  Clump C to  D.  Clump C is visible as the bright spot at the top of the arc marked in Figure 5 and based on it position, Clump D is near the tip of the arc.  Figure 5 shows an arc-like diffuse feature  with increasing velocity with increasing distance from VY CMa  presumably due to an outflow from the star. Thus  we suggest that there is an apparent velocity gradient in the  $^{12}$CO emission towards Clump D with distance from VY CMa. 

\vspace{2mm}

Figure 6 shows the Doppler velocities relative to the star, V$_{rel*}$, of the small separate $^{12}$CO emision spots viewed toward Clump D (see Figure 2) as a function of projected distance from the star. Each point is the centroid position of the local emission in a particular frequency channel.  
Figure 6 illustrates what we see in Figure 5 projected onto the plane of the sky. 
These points fall along a remarkably smooth curve.  The high internal consistency is not 
due to subjective factors, since we employed   objective algorithms and the chosen data points were 
determined by the frequency channels (see Appendix C).

\begin{figure}
\figurenum{6}
\epsscale{0.7}
\plotone{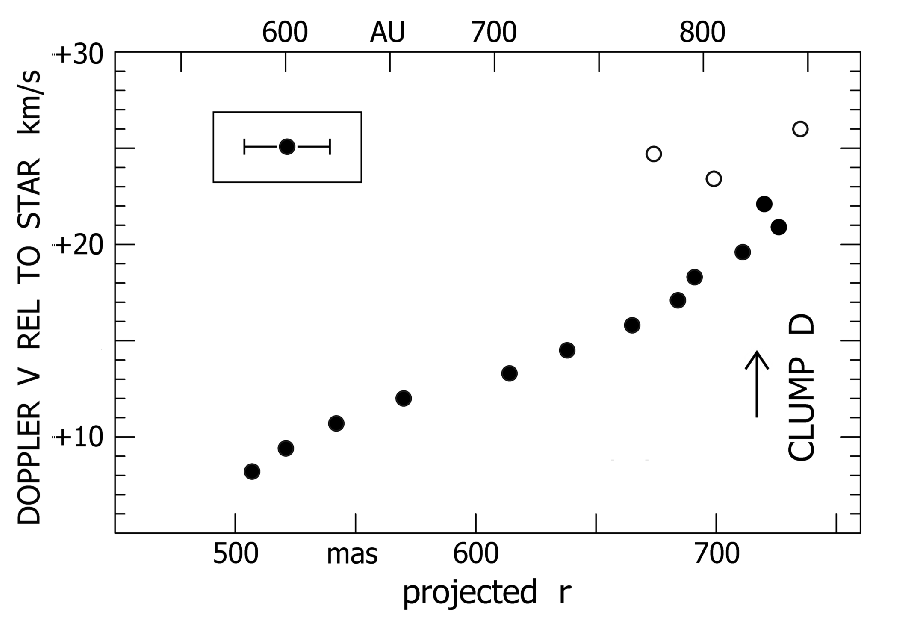}
\caption{The observed line of sight velocity of the $^{12}$CO emission relative to VY CMa toward  Clump D  vs the corresponding projected distance from the star in mas and AU for 15 channels representing the diffuse arc in Figure 5. The open circles show the more diffuse fainter $^{12}$CO emission at the end of the arc in Fig. 5. Each measurement used a sampling parameter of 5 pixels (0\farcs2). The error bar represents the typical uncertainty in the  centroid position.  The main trend agrees with the diffuse arc in Figure 5.}
\end{figure}

\vspace{2mm} 

Together, Figures 5 and 6 demonstrate an outflow of gas with a consistent gradient of increasing
velocity with distance from the star that is not linear, suggesting an arc-like outflow with a maximum 
distance from the star, with no substantial $^{12}$CO emission at farther distances. 
Figure 6 also shows what may be a hook or loop at the very tip of the arc, albeit traced by the more diffuse fainter emission plotted as open circles. Thus the C--D feature may be part of a loop, resembling VY CMa's 
well-known  large stuctures, but smaller and younger.  The $^{12}$CO emission 
may be analogous to the K I  emission observed behind the leading edges of the large expanding 
arcs \citep{RMH2005}.      

\vspace{2mm} 

Clump D is near the tip of the arc and is optically thick in the continuum (see Section 5), so  we do not expect to see emission directly beyond it. It may be part of the outermost loop segment, which is nearly tangential to our line 
of sight  The maximum velocity of CO emission relative to the star at this position is approximately 22 km s$^{-1}$.  This velocity coincides with Clump D's projected angular distance at 0\farcs73 in Tables 1 and 2, marked by an arrow in Figure 6. Consequently, we suggest that this is  the Doppler velocity of the $^{12}$CO gas at or nearest to Clump D and adopt it for Clump D's  line of sight velocity, V$_{rel*}$ with an uncertainty of $\pm$ 3 km s$^{-1}$ based on the range in velocities near Clump D's  position.

\vspace{2mm} 
Clump C is also optically thick. It is located at the top of the diffuse arc in Figure 5, near the position of the star at LSR velocities +25 to +30 km s$^{-1}$ and at $\approx$  0\farcs4 from VY CMa corresponding to its position in the continuum image (Table 1). The line profile (Appendix C) for Clump C shows a small emission peak at  LSR velocity +26 km  s$^{-1}$ which we adopt for its line of sight velocity corresponding to +4 km  s$^{-1}$, V$_{rel*}$ with an uncertainty of  approximately $\pm$ 3 km  s$^{-1}$ based on its position in Figure 5 and the width of the emission peak in the line profile.   

\vspace{2mm} 
Clumps A and B are apparently in an outflow extending to the northeast of VY CMa, similar to the diffuse emission arc including Clumps C and D to the southeast (Figure 1). 
Based on its position angle and projected angular distance from VY CMa, Clump A is very likely associated with the northern extension reported by \citet{Richards} and \citet{OGorman}.  We observe blueshifted CO  emission close to the star beginning at LSR Vel +19 km s$^{-1}$ with a position angle of about 25$^{\circ}$ which we associate with this outflow, and identify Clump A with the $^{12}$CO emission marginally resolved  from the star with an LSR velocity range of +15  to -4 km s$^{-1}$. The blue edge of its emission profile in  Figure 10 is cut off by the absorbing ejecta  (Section 6) at LSR velocities $\lesssim$ -4 km s$^{-1}$. The $^{12}$CO emission in the image cubes agrees with the position of Clump A for the LSR velocity range $\approx$ +8.6 to -1.6 km s$^{-1}$. Adopting the middle of this range at +3.5 km s$^{-1}$, V$_{rel*}$ is -18 $\pm$ 5.0 km s$^{-1}$.  

\vspace{2mm} 

Clump B is more complicated. It is not resolved from the extended circumstellar emission surrounding the star (Figure 2).  It can be seen as an extension on the stellar envelope coinciding with its position in the continuum over a small range of redshifted velocities. Its line profile in Figure 10, however,  shows a very broad emission envelope with two peaks on either side of the star's velocity plus a peak near it and the velocity of Clump C.  Clump B is not distinguishable at velocities blueshifted relative to the star. The emission peak at $\approx$ +8 km s$^{-1}$  overlaps with emission from the NE outflow/Clump A at the same velocity. Clump B cannot be separated from the emission from the circumstellar ejecta coinciding with the 2nd peak near the stellar velocity. So we tentatively adopt the LSR velocity of the redshifted emission peak at +35 km s$^{-1}$. The V$_{rel*}$ is +13 km s$^{-1}$ $\pm$ 7  with the error from the range in the velocities in Table 3.  We consider this to be very uncertain.

\section{Morphology and Ages of the ALMA Clumps} 

The velocities are summarized in Table 4 with the total or expansion velocity relative to VY CMa, the orientation or projection angle ($\theta$) with respect to the plane of the sky, the distance from VY CMa  corrected for the projection, and an estimate of the age or time since ejection measured from the proper motions. 
The reference epoch for the proper motions, 2015, is the reference date for when the clumps were ejected. The age estimates assume no acceleration or deceleration. We discussed the  effect of acceleration or deceleration in \citet{RMH2021} and showed that the effect was minimal, on the order of 5\% to 10\%, less than the errors in the estimated ages. 
\vspace{2mm}

We find that Clump C is nearly in the plane of the sky as suggested by \citet{OGorman} and \citet{Kaminski}. The transverse velocities for Clumps D and A  are rather high for  knots and condensations close to the star, which have typical velocities of 20 to 30 km s$^{-1}$ \citep{RMH2019,RMH2021}, but both have large uncertainties. These uncertainties from the proper motions dominate the error in  the age estimates, the expansion velocities and subsequent derived parameters.  For reference,  the escape velocity for VY CMa, at the stellar surface is $\approx$ 70 km s$^{-1}$ \citep{RMH2021}. 

\begin{deluxetable}{lllllll} 
\tablewidth{0 pt}
\tabletypesize{\footnotesize}
\tablenum{4} 
\tablecaption{Velocities, Distance and Time Since Ejection}
\tablehead{
\colhead{Clump} & 
	\colhead{Trans. Vel.}  &
	\colhead{Doppler Vel.\tablenotemark{a}}  &  
	\colhead{Total Vel.} &
	\colhead{Orient $\theta$} &
        \colhead{Dist.} & 
	\colhead{Age\tablenotemark{b}} \\
\colhead{}          &
\colhead{km s$^{-1}$} &  
\colhead{km s$^{-1}$}  & 
\colhead{km s$^{-1}$} &
\colhead{degrees}  & 
\colhead{AU} &  
\colhead{yrs}   
}
\startdata
	Clump A & 46$\pm$11  &  -18$\pm$5  & 50$\pm$11  &  -21$\pm$7  &  691$\pm$60 &  66$\pm$-12,+19 \\
	Clump B & 18$\pm$9   & +13$\pm$7?\tablenotemark{c} & 23$\pm$8   & 35$\pm$20  & 326$\pm$80   & 69$\pm$-19,+26  \\ 
	Clump C & 20$\pm$2  & +4$\pm$3   &  21$\pm$3 & +11$\pm$8  &  375$\pm$30 & 86$\pm$-7,+8 \\
	Clump D  & 52$\pm$23  & +22$\pm$3  &  57$\pm$23  &  +23$\pm$10  &  895$\pm$90   & 75$\pm$-15,+25  
\enddata
	\tablenotetext{a}{The $^{12}$CO line of sight velocity relative to VY CMa, V$_{rel*}$.} 
\tablenotetext{b}{The reference epoch is 2015 for the ages measured from the proper motions and the projected distance from the star in arcsec.} 
\tablenotemark{c}{Clump B's Doppler velocity is uncertain because of doubtful identification of the 
associated $^{12}$CO emission source, see text.}
\end{deluxetable} 

\begin{figure}
\figurenum{7}
\epsscale{1.0}
\plotone{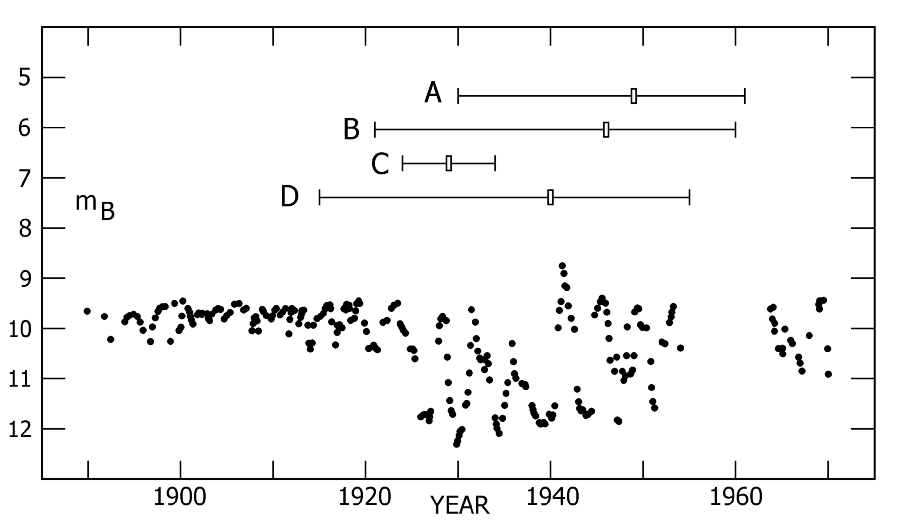}
\caption{the historic light curve for VY CMa from \citet{Robinson} with the estimated ejection 
dates  and possible range of dates from the uncertainty in the expansion velocities for the four clumps.}
\end{figure}

\vspace{2mm}

VY CMa's light curve for the first half of the 20th century is shown in Figure 7 with the probable ejection dates and range. The clumps were all ejected in the early 20th century during VY CMa's very active period 1920 - 1950.   There are seven episodes or deep minima during this period each lasting 2 to 3 years which may correspond to the high mass loss events and accompanying obscuration by dust similar to what was observed  in Betelgeuse but on a larger scale.  Based on the continuum image (Figure 1), it may be tempting to speculate that the four clumps  have a common origin in the same high mass loss episode. Within the 
uncertainties, the clumps could all have been ejected during the same minimum, but this seems unlikely with their different  projection angles and directions they are moving. 

\vspace{2mm}

Figure 1 also shows two major outflows; to the Southeast with Clumps C and D and to the Northeast with Clumps A and B. With their relatively high transverse  velocities Clumps D and A may represent the initial ejection and leading edge of a major  eruption that in the case of the SE outflow includes Clump C,  but their different projections imply different active regions for D and C. 
Based on their motions, we suggest that Clumps A and D are separate events from C.         

\vspace{2mm}  

The largest and brightest, Clump C with the highest quality measurements, has a relatively well-determined ejection time circa 1930\footnote{\citet{RMH2021} estimated a similar ejection time for Clump C adopting the average expansion velocity from the visible knots with the 1999 reference epoch. \citet{Kaminski} showed the same light curve but adopting a higher velocity initially concluded that the clumps did not correspond with the minima.}. Although Clump B's Doppler velocity and spatial orientation are uncertain, Clumps B and C are the closest to VY CMa with  similar proper motions.   This active period might also correspond with the ejections of knots W1 A and W1 B just to the west of VY CMa \citep{RMH2019}. W1 B has an estimated ejection date of circa 1932. If so, their different directions, on opposite sides of VY CMa, would be evidence that surface activity may occur over much of the star during the same short period yielding  separate outflows.   

\vspace{2mm}

Six outflows or ejecta are now identified within VY CMa's $\approx$ 25 year active period beginning about 100 years ago.  Seven minima are observed, thus there were very likely additional ejecta which are obscured or occurred out of our current line of sight. In the next section we discuss the mass lost in these ejections and the energy required. 

\section{Mass Lost and the Mass Loss Mechanism in VY CMa} 

\begin{deluxetable}{lccccll} 
\tabletypesize{\footnotesize}
\tablenum{5} 
\tablecaption{Derived Parameters for the ALMA Clumps}
\tablehead{
\colhead{Clump} & 
\colhead{Flux Density\tablenotemark{a}} &
\colhead{Brightness Temp\tablenotemark{b}}  &
\colhead{Dust Temp}  &
\colhead{Opt. Depth} & 
\colhead{Dust Mass} &
\colhead{Total Mass M$_{\odot}$} \\
\colhead{}          &
\colhead{mJy} &  
\colhead{degrees K}  & 
\colhead{degrees K}  &  
\colhead{$\tau$}   &
\colhead{M$_{\odot}$} &    
\colhead{dust + gas}  
} 
\startdata
	Clump A & 22  &  11  & 352 &  0.03 &  $\approx$ 8$\times$ 10$^{-5}$ & $\approx$ 1.6$\times$ 10$^{-2}$ \\
	Clump B & 33  &  11  & 512   & 0.02 & $\approx$ 1.8$\times$ 10$^{-5}$  & $\approx$ 3.6$\times$ 10$^{-3}$ \\ 
	Clump C & 161  & 243 &  478 & 0.71 &  $>$ 1.2$\times$ 10$^{-4}$ & $>$ 2.4$\times$ 10$^{-2}$ \\
	Clump D  & 5  &  186 &  305 & 0.94 &  $>$ 3$\times$ 10$^{-5}$ & $>$ 6$\times$ 10$^{-3}$
\enddata
\tablenotetext{a}{Continuum emission at 249 GHz, mostly thermal emission by dust grains.}
\tablenotetext{b}{Based on the continuum flux density and apparent projected size of each clump (FWHM). May be under estimated because the clumps are not well resolved.}
\end{deluxetable} 

Measured and derived parameters including the integrated flux density at 249 GHz and the dust mass for each clump are summarized in Table 5. The expected grain or dust temperature, $T_{gr}$, is estimated using the standard approximation 
   $T_{gr} \; \approx \; \zeta \, (R_*/r_{gr})^{1/2} \, T_* $, 
where the efficiency factor $\zeta$ is close to unity for most 
normal  types of grains above $T_{gr}\sim 200$ K 
(see, e.g.,  \citet{Draine}). $r_{gr}$ is the distance of the clump from the star corrected for the projection angle in Table 4, with an adopted temperature of 3500 K \citep{Wittkowski} for VY CMa and radius, $R_*$, $\approx$ 7 AU. A more precise approximation requires information about the grain properties that is not readily available. This approximation assumes that the dust is optically thin. The optical depth $\tau$ is estimated from the standard relation between the observed and dust temperatures\footnote{e$^{-\tau} = 1 - (T_{Obs}/T_{gr})$}. We determined the dust mass  following \citet{OGorman} with the same input parameters. The suggested gas to dust ratio for red supergiants ranges from 100 to a high of 500 for VY CMa \citep{Decin}. For comparison with previous work on VY CMa, we adopt  a  gas to dust ratio of 200 \citep{Mauron} for the total mass lost.  The mass estimates, especially for Clumps C and D, are lower bounds since the calculation uses the dust temperature for an optically thin case. 

\vspace{2mm}  

Clumps C and A have our highest mass estimates. Our result for Clump C  is comparable to that from \citet{OGorman}, but is less than the value from \citet{Vlemmings}.  The masses for the other two clumps are typical for other knots in VY CMa with dust masses based primarily on their infrared fluxes, see Table 1 in \citet{RHJ}.  Clumps C and A may be representative of the more massive outflows or ejecta in VY CMa. The energies required though is quite modest for VYCMa with a luminosity of 3 $\times$ 10$^{5}$ L$_{\odot}$. For example, Clump C's kinetic energy is a little more than 10$^{44}$ ergs, assuming its outflow velocity of 21 km s$^{-1}$,   equivalent to VY CMa's luminosity in less than a day.

\vspace{2mm}  

The total mass shed during this active period by the four ALMA knots is $\ge$ 0.05 M$_{\odot}$. Assuming 10$^{-2}$ M$_{\odot}$  for knots W1 A and  W1 B, the estimated mass lost from the observed knots and clumps  is at least  0.07  M$_{\odot}$ yielding an effective mass loss rate of $\approx$  10$^{-3}$  M$_{\odot}$ yr$^{-1}$ or more  in 25 to 30 yrs. The mass lost in these discrete episodes not only dominates VY CMa's recent mass loss history \citep{RMH2021,RHJ}, but explains its measured high mass loss rate of 4  to 6 $\times$ 10$^{-4}$ M$_{\odot}$ yr$^{-1}$ \citep{Danchi,Shenoy}.  The mass loss estimate from the recent dimming of Betelgeuse \citep{Montarges} together with the mass in its circumstellar condensations \citep{Kervella09,Kervella11}, also contribute significantly to its measured mass loss rates \citep{RHJ}.  Betelgeuse and the similar recent dimming of the high luminosity K-type supergiant RW Cep \citep{Jones23,Gies} are increasing evidence that high mass surface outflows are more common in RSGs and a major contributor to their mass loss.

\vspace{2mm} 

The primary questions concern the mass loss mechanism  for the high mass loss episodes.  The standard methods for mass loss that work well for red giants and AGB stars, e.g. radiation pressure on grains, pulsation and convection, are not adequate for RSGs and cannot account for the elevation of the material to the dust formation zones \citep{Arroyo13,Arroyo15,Climent} in the very dusty, high luminosity RSGs like VY CMa. 
\vspace{2mm} 

The dynamical timescale for VY CMa is about three years, comparable to the timescales for a non-radial instability or a magnetic/convective event similar to Coronal Mass Ejections (CME).
The presence of magnetic fields in VY CMa is supported by the circular polarization and Zeeman effect observed in the OH, H$_{2}$O, and SiO masers in its ejecta \citep{Vlemmings02,Richter,Vlemmings}.  \citet{Vlemmings} reported polarized dust emission from magnetically aligned grains on sub-arcsec scales close to the star.  \citet{RHJ} estimated a surface field of 500G based on the magnetic field strengths from the maser emission in the ejecta. They also showed that scaling the most energetic Solar CME's to the kinetic energies of the knots in VY CMa, would require a 1000 G field similar to the above estimate.

\vspace{2mm}

Fundamental clues to VY CMa's mass loss can be seen in 
 the morphology of its ejecta.  In this regard, stellar 
 outflows can be categorized by their acceleration mechanisms:    
 (1) Supernova outflows are driven by blast waves. They generally 
 produce numerous filaments, often with roughly spheroidal 
 symmetry.  Nova remnants may also fit into this category.  
 (2) Giant eruptions are driven by continuum radiation pressure 
 \citep{Davidson}. The only well-resolved example, $\eta$ Car, 
 exhibits fossil turbulence cells that visually resemble 
 volcano eruptions and some other terrestrial-scale explosions.  
 (3) Planetary nebulae are driven by dynamical instabilities. 
 Their filamentary systems often include small mass 
 condensations and large-scale bipolar or multipolar 
 symmetries.  
 (4) Solar or stellar activity is driven by interactions 
 of convection and plasma physics with magnetic fields.  
 The outbursts are spatially localized rather than global, 
 with loops and arcs.     

\vspace{2mm} 

The morphology of VY CMa's ejecta is most similar to stellar activity.
A lack of global symmetry 
 is obvious, and each loop or arc (see \citet{RMH2005,RMH2007} 
 and Fig.\ 9 below) may be a fossilized remnant of initial 
 magnetic  confinement of outflowing material.  The same 
 process   may also create localized mass concentrations 
 like the knots or clumps.  The major question is whether 
 MHD-like processes can liberate enough energy for the 
 outburst,  in addition to confining it.

   \vspace{2mm}

\begin{figure}
\figurenum{8}
\epsscale{0.8}
\plotone{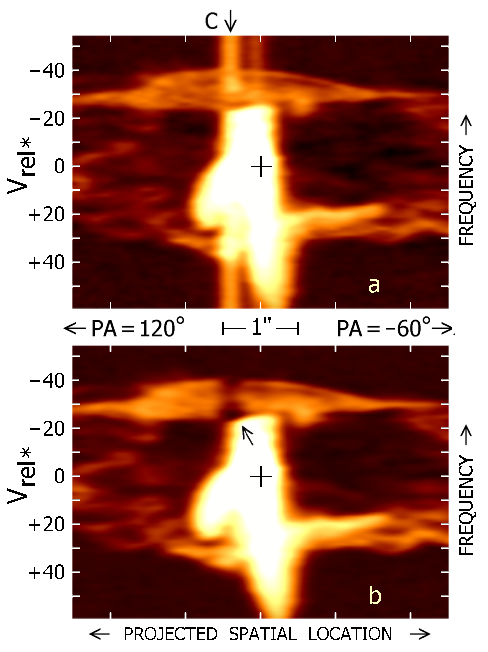}
\caption{ An oblique plane in the image cube, similar to Figure 5. 
     The vertical scale is frequency, represented by the $^{12}$CO   
     Doppler velocity relative to the star.  The horizontal axis  
     is spatial position along a projected line oriented at 
     position angle 120$^{\circ}$, cf.\ Fig.\ 4.  The star's 
     position is marked by '+'.     
     (a) Total observed intensity, $^{12}$CO emission plus continuum.   
     The prominent vertical line just left of center is continuum 
     emission from Clump C.  (b) Same image after the formal 
     continuum subtraction process.  An arrow marks the invalid 
     dark spot shown in Figure 9 and explained in the text. } 
\end{figure} 

\section{Unexpected phenomena in the outer ejecta}  

Figure 5 in Section 3 was designed to illustrate the Clump C--D 
structure, but it also reveals remarkable larger-scale details   
that were not recognized earlier.  

     \vspace{2mm}  

Figure 8 is similar to Fig.\ 5, but has brightness and contrast 
parameters that show the outer ejecta more clearly.  Again the 
horizontal scale represents projected spatial location along a line 
with position angle $120^{\circ}$ and $-60^{\circ}$ (roughly ESE and WNW), 
centered at the star.  The vertical scale is frequency expressed as 
$^{12}$CO Doppler velocity relative to the star -- e.g., negative 
velocities at the top of the figure represent ejecta moving toward us.    

\vspace{2mm}

Unlike most other figures in this paper, Fig.\ 8a includes the total 
observed $^{12}$CO emission plus continuum;  continuum has not been 
subtracted.  Clump C, the strongest continuum source, produces a 
conspicuous vertical feature about 0.4 arcsec to the left of the star.  
The bright vertical structure near the center of the figure is 
mostly $^{12}$CO emission in the inner material with $r < 600$ AU, 
ejected less than 200 years ago.  Horizontal cirrus-like features 
around $V_{rel*} \, \sim \, \pm 30$ km s$^{-1}$, on the other hand, 
represent older ejecta at $r \, \sim$  3000 to 8000 AU,  
with ages $\gtrsim$ 800 y.  This molecular emission was reported 
earlier by \citet{Ziurys} and in Paper I, while \citet{Shenoy} 
described IR emission 
from the associated dust.  $^{12}$CO images at various frequencies 
confirm that the pattern of inner- vs.\ outer ejecta is broadly similar 
toward other position angles.  

   \vspace{2mm}

The upper half of Fig.\ 8a reveals a fact that was not recognized 
earlier:  the outer ejecta are opaque in a very narrow range   
of frequencies corresponding to $^{12}$CO $V_{rel*} \approx -28$ 
km s$^{-1}$.  Pictorially the cirrus-like layer forms a conspicuous  
ceiling to the inner-ejecta emission, and the latter is 
undetectable in several of the frequency channels.  At those frequencies 
the outer material is optically thick, with brightness temperatures in 
the range 20--40 K along the lines of sight that we discuss here.  
This is remarkable, and arguably surprising, for several reasons.  
(1) In most cases of old, strongly inhomogeneous stellar ejecta, 
some radiation escapes through interstices or gaps between the 
filaments or condensations even if each condensation is optically 
thick.   But that is evidently untrue here.  
(2) The transition is remarkably abrupt in frequency space;  
for instance, continuum from Clump C is detectable at 
$V_{rel*} = -22$ km s$^{-1}$ but undetectable at $-25$ km s$^{-1}$.  
(3) The velocity range of opaque material is only about 8 km s$^{-1}$, 
even though the outer ejecta are extremely non-spherical.   
We cannot explore the physical implications of a fully-opaque 
molecular shell here, because that is beyond the scope of this 
paper.  At present the main point is that one must be aware 
that it affects our view of the inner ejecta.

    \vspace{2mm} 

Moreover, this case provides a dramatic example of a strange 
illusion generated by the standard  data reduction described in Paper I and here in Appendix A.  Thoughout this 
paper, we have attempted to remove the continuum emission in order 
to focus on $^{12}$CO emission. Generally a renormalized 249 GHz 
    continuum image  (Fig.\ 1) is subtracted from the data image 
for each frequency channel.        
    This procedure is satisfactory at frequencies where the 
    intervening material is transparent.  However, it becomes  
    invalid near $V_{rel*} \approx -28$ km s$^{-1}$  
    because the inner continuum sources are entirely hidden 
    at those frequencies by $^{12}$CO absorption and emission 
    in the outer ejecta as described above.  Moreover, Clump C 
    dominates in the subtracted continuum image. Figure 9 shows  
the peculiar result.  Figure 9a, like 8a, represents the total 
observed data without continuum subtraction;  it is 
almost entirely $^{12}$CO emission with $T_b \sim 30$ K 
produced in the opaque layer, with  no perceptible contribution 
by Clump C.  Figure 9b is the image after continuum subtraction;  
it shows a very conspicuous dark spot which is merely a 
negative image of Clump C.  \citet{Kaminski} assumed that this 
is a real feature, which would require strange attributes 
(e.g., a mismatch between size and internal velocity dispersion).  
In fact, as stated above, it is an artifact caused by unsuitable 
continuum subtraction.  It can also be seen in Figure 8b, marked by an 
arrow.

\begin{figure}
\figurenum{9}
\epsscale{0.9}
\plotone{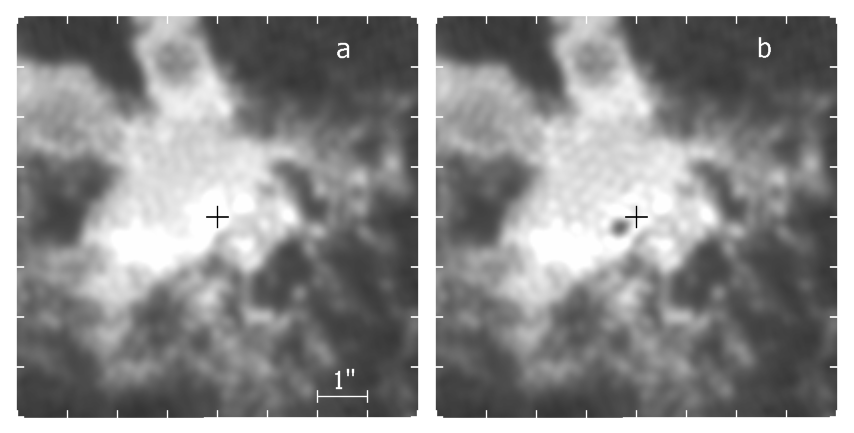}
  \caption{ $^{12}$CO spatial images at Doppler velocity 
   $V_{rel*}$ -27 km s$^{-1}$. At this frequency the outer ejecta are 
   optically thick, hiding the inner ejecta. 
   (a) Observed intensity, almost entirely $^{12}$CO emission. 
   (b) Result of the formal continuum subtraction procedure 
   which is valid at most other frequencies.  The prominent 
   central dark spot is an illusory negative image of 
   continuum from Clump C.  In both images, the mottled   
   mesh pattern is an artifact of the interferometry 
   reduction process. }  
\end{figure}

\vspace{2mm}

\section{Comments on the Evolutionary State of VY CMa}

VY CMa's observed properties suggest a very evolved red supergiant that  may  be near the end of the RSG stage. It is one of most luminous known RSGs with a corresponding  initial mass of $\approx$ 30 to  35 M$_{\odot}$, significantly above the 18M$_{\odot}$ upper mass for the progenitors of the Type IIP supernovae \citep{Smartt}. Thus its final fate may be to either directly collapse to a black hole or evolve back to warmer temperatures before the terminal explosion. Stellar structure models \citep{Eggenberger} show that the latter requires sufficient mass loss to increase the ratio of the He/C core relative to the total mass to send the star on a ``blue loop'' across the H-R Diagram.

\vspace{2mm} 

Our observations of the CO emission from the outer ejecta discussed in the previous section, reveal diffuse filamentary ejecta expanding at 30 to 40 km s$^{-1}$ relative to the star at distances of  5 to 6$\arcsec$ from VY CMa \citep{Ziurys,Singh}. At about 7000 AU from the star, the expansion age of the outflow is 800 to 1100 years.  The CO emission appears to surround the more complex inner ejecta including the large arcs in the HST images. The prominent Arc 1 is the oldest with an expansion age  of about 800 years \citep{RMH2007}. In an independent study based on a radiative transfer model of the CO emission, \citet{Decin}  concluded that a high mass loss phase in VY CMa began  about 1100 years ago. Furthermore, the 37$\mu$m image of the cold dust \citep{Shenoy} shows that there is no cold dust beyond a radius of 10\arcsec, corresponding to an ejection age of about 1400 years. We thus conclude that VY CMa entered its presently observed active phase with relatively frequent massive outflows about 1200 years ago.  

\vspace{2mm} 

We have no explanation for the onset of its enhanced surface activity which may have  been stimulated by  a change in the interior or more likely a change in the structure of the convective layers. Of course, VY CMa is not alone. Surface outflows have  been observed in Betelgeuse and RW Cep, although the record for VY CMa is remakable.  

\vspace{2mm}

VY CMa also has a unique rich and peculiar chemistry. Twenty-five molecules have been identified in its ejecta including molecules of carbon and silicon. The only comparable star, is the similar luminous RSG, NML Cyg \citep{Singh2} with 21  molecules in common with VY CMa.  
In Paper I, we reported on preliminary $^{12}$C/$^{13}$C ratios of 22 to 38 in 
various structures in the ejecta  These ratios are significantly higher than measured in oxygen-rich red giants and supergiants and may be indicators of additional dredge-up perhaps related to VY CMa's surface activity. Our future work will include abundance measurements of the additional molecules observed in the program (Paper I). Their association with 
separate outflows, arcs, and clumps at different locations and with expansion ages may provide more clues to VY CMa's current state and fate. Another possibility, not often discusssed, is whether VY CMa may be a second generation red supergiant. Similar to lower mass stars, it may have evolved back to warmer temperatures,  has now returned, and is in a very short high mass loss state prior to core collapse to a black hole.


\appendix 

\section{Parameters for the 2021 ALMA Observations}

The parameters for the continuumm and CO images used in this paper are summarized in Table A1.  The beam properties are the major and minor axes and position angle. The rms noise $\sigma_{\mathrm{rms}}$ is measured within the region of full sensitivity but away from bright emission. The MRS is the maximum recoverable scale of emission which can be imaged accurately. Both cubes were made to cover 36'' on a side, out to $<0.3$ of the full primary beam sensitivity.

\vspace{2mm} 

The 2021 data are from the observations described in Paper I, (covering a larger range of frequencies and angular scales). We started from the data delivered from the ALMA Science Archive with all instrumental calibration as well as bandpass, flux scale and phase referencing corrections applied.  We split out the VY CMa data, which were adjusted to constant velocity in the VLSR convention in the direction of VYCMa. For each tuning and array configuration, we selected the line-free channels to use these for phase and then amplitude self-calibration.   We applied the calibration to all channels.  The resolution of the more extended-configuration data varies slightly between tunings and we chose the highest-resolution continuum image for this analysis.  For each spectral window, the line-free continuum was substracted from the calibrated visibility data using a first-order linear fit for the image cubes. The image cubes were also made from the unsubtracted data, using the same parameters, for the 12CO, 13CO and HCN lines. 
The 2015 data are described by \citet{Kaminski}.

\begin{deluxetable}{lcccccccc}
\tablewidth{0 pt}
\tabletypesize{\scriptsize}
\tablenum{A1}
\tablecaption{Characteristics of the Continuum Images and $J$=2-1  CO Image Cubes}
\tablehead{
\colhead{Transition} & 
\colhead{Date} &
\colhead{Center $\nu$} &
\colhead{No. chan} &
\colhead{Chan. width} &
\colhead{V$_{\mathrm{LSR-min}}$ to V$_{\mathrm{LSR-max}}$ } &
\colhead{MRS}  &
\colhead{Restoring beam} &  
\colhead{$\sigma_{\mathrm{rms}}$}   \\
\colhead{}        &
\colhead{yyyymmdd}  &    
\colhead{GHz}     &  
\colhead{}     &
\colhead{km s$^{-1}$}  &
\colhead{km s$^{-1}$}  &
\colhead{arcsec}  &
\colhead{mas$\times$mas deg} & 
\colhead{mJy}  
}
\startdata 
  Continuum &20150927 &230.314     &  --     &  --        & --                  &  1''.6  &  175$\times$132 70$^{\circ}$&  0.1\\
  Continuum &20211209 &249.339     & --      &  --       &  --                  &  3''.4  &  196$\times$180 39$^{\circ}$&  0.05\\
  $^{13}$CO &20211216& 220.379      & 144     &  1.33     &--68.58 to +121.38   &  3''.4  &  257$\times$241 11$^{\circ}$&  1.0\\
  $^{12}$CO &20211216& 230.513      & 158     &  1.27     &--58.74 to +140.63   &  3''.4  &  265$\times$251 11$^{\circ}$&  0.9
\enddata  
\end{deluxetable}

\section{The Fitting Method for the Proper Motion Measurements for the Four Clumps} 

The 2021 continuum image (Paper I) used for measuring the proper motion has resolution 0\farcs20 $\times$ 0\farcs18, rms 0.04 mJy, central frequency 249.34 GHz. The  absolute astrometry for the 2015 data \citep{Kaminski}  may be worse as at that time antenna positions were less well-determined but our
 comparison uses only relative positions. The 2015 continuum image  used for measuring proper motions has resolution 0\farcs17 $\times$ 0\farcs13, rms 0.1 mJy, central frequency 230.31 GHz.  

\vspace{2mm} 

We used the CASA task imfit for fitting the continuum peaks of the four Clumps A, B, C and D plus VY CMa, see Figure 1.  We treated both the 2015 and 2021 data similarly to ensure comparability.  Estimates were first made of the individual peak positions and flux densities using the CASA viewer. 2D Gaussian compoments were then fit to all five peaks simultaneously, using input estimates for the peak flux densities, positions, angular sizes and position angles. The estimates were refined iteratively to reduce the residuals. The residual after fitting to the star only was used to confirm the initial  position estimate for B. As the accuracy improved, the peak flux estimates were fixed (starting with the brightest components), leaving the positions and sizes (and thus integrated fluxes) as free parameters. For the 2015 data, once good (stable) position estimates had been obtained for the brightest components these also were fixed in order to avoid the weaker
components having unphysically large sizes. Simultaneous fitting should ensure that all the peaks are fit with comparable accuracy and minimise leakage or conversely over-subtraction between components.  The relative position uncertainties given in Table 1 due to phase noise are approximately (synthesised beam)/(S/N), for moderately extended array configurations (ALMA memo 620). We adopt  the same errors in both coordinates as the synthesised beams are close to circular and have a small elongation, at intermediate angles. These were propagated for the measurements in Table 2. 

\vspace{2mm} 

An additional source of uncertainty arises because the distributions of emission for clumps A, B, C and D are not necessarily Gaussian, and  they may have sub-structure which evolves in brightness during the outflow.  The apparent fitted, deconvolved sizes varied between ~2 synthesised beams and being unresolved but we do not consider the Gaussian approximations good enough to analyse the sizes.  We did check as described that 
in all cases apparent radii of adjacent clumps were significantly less than half the angular separation, so any potential overlap is at too low a level to bias the fit significantly.

\vspace{2mm} 

After subtracting the fits images, the residual images have rms 0.4 and 0.2 mJy for the 2015 and 2021 data, respectively. Positions in Table 1 are given to an extra decimal place to avoid rounding errors in calculating offsets.

\section{The Line Profiles for the Four Clumps and Accuracy of the Doppler Velocities}

\begin{figure}
\figurenum{10}
\epsscale{0.8}
\plotone{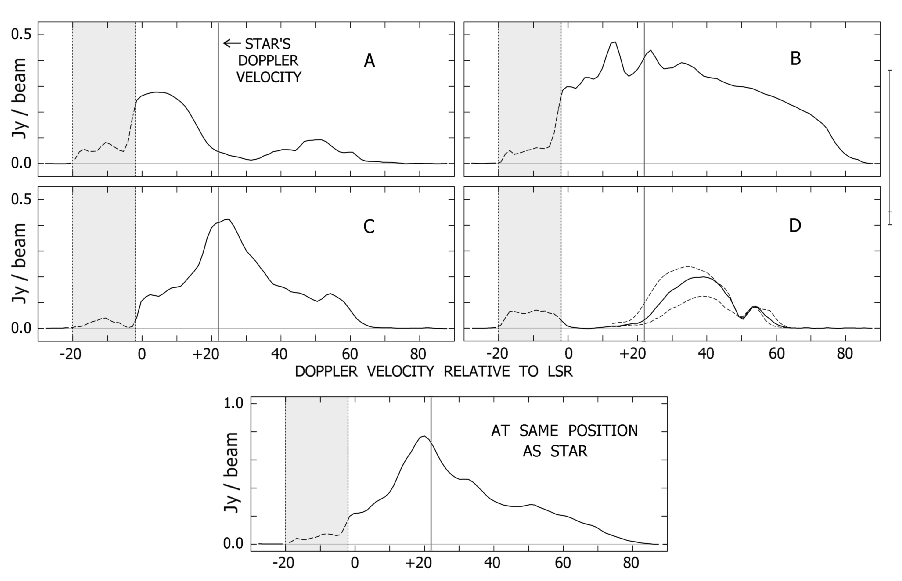}
\caption{The continuum subtracted $^{12}$CO line profiles at the projected 
centers of Clumps A, B, C, and D with a 1 pixel circular aperture.    
In the range -20 to -2 km s$^{-1}$ on the left side, the shaded areas 
represent an opaque outlying shell which conceals the inner ejecta 
(Figure 8 and Section 6).  As explained in the text, the two dashed 
curves represent locations 100 mas NW and SE of the center of Clump D  
(NW is brighter, cf.\ Figure 5).  A fifth panel shows the intensity along 
the line of sight to the star, but this emission arises in circumstellar ejecta rather than the star. }
\end{figure}

   \vspace{1mm}  

Figure 10 shows velocity profiles along lines of sight through Clumps 
A, B, C, D, and the central star.  For each clump the plotted quantity 
is specific intensity at the pixel nearest to the adopted clump-center 
position;  it adequately represents the clump because 
pixel-to-pixel local noise is negligible in the processed ALMA data.  
Gaussian statistics cannot be applied to errors in the velocities 
of the peaks in Figure 10, because those errors are mainly systematic -- 
especially due to spatial confusion with unrelated 
emission within 100 mas.  For each measurement one must inspect the 
spatial images for several frequency channels.  Error values quoted 
for Doppler velocities in Table 4 represent, informally, a confidence 
level roughly equivalent to $\sigma$ used  in formal statistics.  
In cases lacking additional information, the realistic uncertainty 
of a peak in these data tends to be roughly ${\pm}$ FWHM/5.  

   \vspace{2mm} 

A more elaborate error analysis is not worthwhile, because most of 
the calculated quantities in Table 4 are insensitive to Dopper velocity. 
For example, suppose that Clump D's Doppler velocity has an unexpectedly 
large error of 5 km s$^{-1}$ (cf.\ Figure 6).  That would alter the 
total expansion velocity and distance from the star by less than 
4\%, and would change $\theta$ by less than $5^{\circ}$.  
The ejection times were estimated without reference to Doppler 
velocities (Section 4).  

    \vspace{2mm}  

Each clump A, B, C, D has individual characteristics.  
For instance, Figure 10 shows that one wing of Clump A's line profile 
($\lesssim -2$ km s$^{-1}$) is concealed by the opaque shell of outer 
material described in Section 6.  Hence we used only a small velocity 
interval to estimate its peak.  

    \vspace{2mm}  

Clump B is ambiguous with 200 mas spatial resolution, 
because other, nearby complex emission from the extended circumstellar ejecta from VY CMa that i
s bright is CO emission, plus the outflow to the northeast. Inspecting the images, we cannot con
fidently identify 
a unique $^{12}$CO spatial match to the continuum image 
of Clump B at any particular frequency shown in Figure 10.  
It is visible in the continuum image, but identification in CO emission is
uncertain. See the discussion in Section 3.1. Hence we cannot reliably measure its 
Doppler velocity.  

    \vspace{2mm}  

Clump D represents the outer terminus of an elongated structure 
(Figures 4--6), not the center of an isolated feature.  Hence 
the limited spatial resolution has an asymmetric effect:  the line 
profile is contaminated by lower-velocity material to the northwest, 
while there is much less compensating higher-velocity emission  
within the p.s.f.  This fact is indicated by two dashed curves in 
Figure 10, showing the profiles at distances of 100 mas in opposite 
directions from the adopted center of D -- specifically, toward 
position angles $-60^{\circ}$ and $120^{\circ}$ aligned with  
the C--D feature (Figure 4).  Due to this asymmetric 
contamination, the peak of D's profile in Figure 10 corresponds to 
a Doppler velocity that is several km s$^{-1}$ lower than the 
value listed in Table 4. 

   \vspace{2mm}  

In these circumstances we measured the C--D feature as $r(V) =$ 
spatial positions at well-defined Doppler velocities, 
rather than $V(r) =$ velocities at given positions.  
Most of the resulting points in Figure 6 fall along a curve that 
is remarkably smooth, considering that the measurement process 
was entirely objective.  (For each ALMA frequency channel a 
specific peak-location algorithm was employed, see Section 6.)  
The horizontal error bar in Figure 6, $\pm$ 20 mas, was  informally 
chosen to be about 50\% larger than the r.m.s.\ deviation 
in a linear fit to the 12 main data points.  Along the 
main trend line it implies an r.m.s.\ velocity error less than 
1 km s$^{-1}$.  Of course this refers only to differences among 
the plotted points, and larger systematic  effects may alter the 
spatial scale.  But the main point here is that informal 
uncertainties of $\pm$ 3 km s$^{-1}$ are quite reasonable for 
the Doppler velocities in Table 4.

\acknowledgments

This paper makes use of the following ALMA data:ADS/JAO.ALMA$\#$2021.1.01054.S and 
ADS/JAO.ALMA$\#$2013.1.00156.S. ALMA is a partnership of ESO (represent-
ing its member states), NSF (USA), and NINS (Japan), together with NRC
(Canada) and NSC and ASIAA (Taiwan) and KASI (Republic of Korea),
in cooperation with the Republic of Chile. The Joint ALMA Observatory is
operated by ESO, AUI/NRAO, and NAOJ. The National Radio Astronomy
Observatory is a facility of the National Science Foundation operated under
cooperative agreement by Associated Universities, Inc. 
This research is supported by NSF Grants AST-1907910 and AST-2307305.


\begin{thebibliography}{}

\bibitem[Anugu et al(2023)]{Gies}Anugu, N., Baron, F., Gies, D. R. et al. 
2023, \aj, 166, 78A

\bibitem[Arroyo-Torres et al.(2013)]{Arroyo13} Arroyo-Torres, B.,  Wittkowski, M., Marcaide, J. M. \&  Hauschildt, P. H. 2013, \aap, 554, A76  

\bibitem[Arroyo-Torres et al.(2015)]{Arroyo15}Arroyo-Torres, B., Wittkowski, M., Chiavassa, A. et al. 2015, \aap, 575, A50  

\bibitem[Asaki et al.(2020)]{Asaki}Asaki, Y., et al. 2020, \apjs, 247, 23

\bibitem[Choi et al.(2008)]{Choi}Choi, Y. K., Hirota, T., Honma, M. et al., 2008, \pasj, 60, 1007  

\bibitem[Climent et al.(2020)]{Climent}Climent, J. B., Wittkowski, M., Chiavassa, A. et al. \aap, 635, A160  

\bibitem[Danchi et al.(1994)]{Danchi}Danchi, W. C, Bester, M., Degiacomi, L. J.,  \& Townes, C. H. 1994, \aj, 107, 1469 

\bibitem[Davidson(2020)]{Davidson}Davidson, K. 2020, {\it Galaxies}, 8, issue 1, 10  

\bibitem[Decin et al.(2006)]{Decin}Decin, L., Hony, S., de Koter,A. et al. 2006, \aap, 456, 549

\bibitem[Draine(2011)]{Draine}Draine, B.T. Physics of the Interstellar Medium, Princeton University Press, pp 248--258. 

\bibitem[Eggenberger et al.(2021)]{Eggenberger}Eggenberger, P., Ekstrom, S.,Georgy, C. et al. 2021, \aap,, 652,A127 

\bibitem[Humphreys et al.(2005)]{RMH2005}Humphreys, R. M., Davidson, K., Ruch, G., \& Wallerstein, G.  2005, \aj, 129, 492

\bibitem[Humphreys et al.(2007)]{RMH2007}Humphreys, R. M., Helton, L. A., \&  Jones, T. J.  2007, \aj, 133, 2716  

\bibitem[Humphreys et al.(2019)]{RMH2019}Humphreys, R. M., Ziurys, L. M., Bernal, J. J., et al. 2019, \apjl, 874L, 26H

\bibitem[Humphreys et al.(2021)]{RMH2021}Humphreys, R. M., Davidson, K., Richards, A.M.S., et al.  2021, \aj, 161, Issue 3
, Id 98 

\bibitem[Humphreys \& Jones(2022)]{RHJ}Humphreys, R. M. \& Jones, T. J. 2022, \aj, 163, 103H

\bibitem[Jones et al(2023)]{Jones23}Jones, T.J., Shenoy, D., \& Humphreys, R. M. 2023, RNAAS, 7, 92J

\bibitem[Kaminski(2019)]{Kaminski}Kaminski, Th. 2019, \aap, 627, 114

\bibitem[Kervella et al.(2009)]{Kervella09}Kervella. P., Verhoelst, T., Ridgeway, S.T., Perrin, G., Lacour, S. et al. 2009, \aap, 504, issue 1

\bibitem[Kervella et al.(2011)]{Kervella11}Kervella, P., Perrin, G, Chiavassa, A., Ridgeway, S. T., Cami, J., et al. 2011, \aap, 531, A117

\bibitem[Mauron \& Josselin(2011)]{Mauron}Mauron, P. \& Josselin, E. 2011, \aap, 526, A156 

\bibitem[Montarg\`{e}s et al.(2021)]{Montarges}Montarg\`{e}s, M., Cannon, E., Lagadec, E., et al. 2021, Nature, 594, 365 

\bibitem[O'Gorman et al.(2015)]{OGorman} O'Gorman, E., Vlemmings, W., Richards, 
A. M. S., et al.  2015, \aap, 573, L1 

\bibitem[Richards et al.(2014)]{Richards}Richards, A. M. S., Impellizzeri, C. M. V., Humphreys, E. M., et al. 2014, \aap, 572, L9 

\bibitem[Richter et al.(2016)]{Richter}Richter, L., Kemball, A., \& Jonas, J. 2016, \mnras, 461, 2309 

\bibitem[Robinson(1970)]{Robinson}Robinson, L. 1970, Inf. Bull. Var. Stars, 465 

\bibitem[Shenoy et al.(2016)]{Shenoy}Shenoy, D. P., Humphreys, R. M. \& Jones, T. J., et al. 2016, \aj, 151, 51   


\bibitem[Singh et al.(2022)]{Singh2}Singh, A. P., Edwards, J. L. \& Ziurys, L, M. 2022,  \aj, 164, 230 

\bibitem[Singh et al.(2023)]{Singh}Singh, A. P., Richards, A. M. S., Humphreys, R. M., Decin, L., \& Ziurys, L. M, 2023, \apjl, in press  

\bibitem[Smartt(2015)]{Smartt}Smartt, S. J. 2015, \pasa,  32, e016 


\bibitem[Vlemmings et al.(2002)]{Vlemmings02}Vlemmings, W. H. T., van Langevelde, H. J., \& Diamond, P. J. 2002, A \&A, 394, 589 

\bibitem[Vlemmings et al.(2017)]{Vlemmings}Vlemmings, W. H. T., Khouri, T., Martí-Vidal, I. et al. 2017, \aap, 603, 92 

\bibitem[Wittkowski et al.(2012)]{Wittkowski}Wittkowski, M., Hauschildt, P. H., Arroyo-Torres, B., \& Marcaide, J. M. 2012,  \aap, 540, L12  

\bibitem[Zhang et al.(2017)]{Zhang}Zhang, B., Reid, M. J., Menten, K. M., \& Zheng, X. W. 2012, \apj, 744, 23  

\bibitem[Ziurys et al.(2007)]{Ziurys}Ziurys, L. M., Milam, S. N., Apponi, A. J., \& Woolf, N. J. 2007, Nature, 447, 1094  

\end{thebibliography}
\end{document}